# Low voltage local strain enhanced switching of magnetic tunnel junctions

*Suyogya Karki[1,2], Jaesuk Kwon[1,2], Joe Davies[3], Raisa Fabiha[4], Vivian Rogers[1,2], Thomas Leonard[1,2], Supriyo Bandyopadhyay[4], and Jean Anne C. Incorvia\*[1,2]*

[1]Department of Electrical and Computer Engineering, The University of Texas at Austin, 2501 Speedway, Austin, TX 78712 USA

[2]Microelectronics Research Center, The University of Texas at Austin, 10100 Harry Ransom Trail, Austin, TX 78758 USA

[3]NVE Corporation, 11409 Valley View Rd, Eden Prairie, MN 55344 USA

[4]Department of Electrical and Computer Engineering, Virginia Commonwealth University, 601 W Main St, Richmond, VA 23284 USA

**Abstract**

Strain-controlled modulation of the magnetic switching behavior in magnetic tunnel junctions (MTJs) could provide the energy efficiency needed to accelerate the use of MTJs in memory, logic, and neuromorphic computing, as well as an additional way to tune MTJ properties for these applications. State-of-the-art CoFeB-MgO based MTJs still require too high voltages to alter their magnetic switching behavior with strain. In this study, we demonstrate strain-enhanced field switching of nanoscale MTJs through electric field control via voltage applied across local gates. The results show that record-low voltage down to 200 mV can be used to control the switching field of the MTJ through enhancing the magnetic anisotropy, and that tunnel magnetoresistance is linearly enhanced with voltage through straining the crystal structure of the tunnel barrier. These findings underscore the potential of electric field manipulation and strain engineering as effective strategies for tailoring the properties and functionality of nanoscale MTJs.

**Introduction**

The magnetic tunnel junction (MTJ) is a fundamental building block of spintronic devices with application in memory[1], logic[2], neuromorphic computing[3,4], and probabilistic computing[5,6]. MTJs are trilayer structures composed of two ferromagnetic electrodes separated by a thin insulating barrier. The MTJ is characterized by the tunnel magnetoresistance $TMR = \frac{R_{AP}-R_P}{R_P} \times 100\%$ resulting from the resistance difference between antiparallel (*AP*) magnetized ferromagnetic electrodes (higher resistance $R_{AP}$) and parallel (*P*) magnetized electrodes (lower resistance $R_P$). Typically, the ferromagnetic layer in an MTJ device is switched using magnetic field, spin transfer torque (STT), or spin orbit torque[7]. However, these switching mechanisms require high current density, resulting in high energy dissipation in MTJs. This motivates research in voltage-controlled switching of magnetization, particularly through techniques such as voltage-controlled magnetic anisotropy (VCMA)[8], and voltage-induced strain for energy efficient switching of MTJs[9,10]. Among them, voltage-induced strain is theoretically a highly energy-efficient method for switching magnetization, predicted to be dissipating 1000 times lower energy compared to STT and at least 10 times lower energy compared to VCMA switching of MTJs[11,12]. Nearer-term, strain can also be used in conjunction with other switching methods to increase the tunability of the MTJ resistance states, for example for use in probabilistic computing[13].

Strain enables control of magnetic anisotropy by coupling a magnetostrictive, ferromagnetic layer with a piezoelectric substrate[14]. When strain is applied, it induces magnetic anisotropy, leading to a change in magnetization direction known as inverse magnetostriction or the Villari effect[15]. To achieve energy-efficient strain-based switching, studying materials with larger high magnetostriction coefficient ($\lambda_s$) is crucial, as it reduces the strain requirement, lowering the voltage needed. Previous studies have demonstrated a high $\lambda_s$ in FeGa[16] and Terfenol-D[17] sputter-deposited films. However, FeGa is thermally unstable due to gallium's low melting point of 26 °C, making it prone to diffusion during post-deposition ultra-high vacuum (UHV) annealing, which is essential for high *TMR* MTJs. Meanwhile, Terfenol-D, being a quaternary alloy with multiple phases, presents challenges for the sputter deposition of several nanometer thin layers[17]. Thus, it is difficult to integrate these high magnetostrictive materials into MTJs and requires fine material engineering techniques. An alternative approach for achieving efficient switching is to maximize strain through electric fields in state-of-art CoFeB MTJs[18]. Previous studies have explored controlling the magnetization of the CoFeB layer using anisotropic strain induced from electric field on the piezoelectric substrate Pb $(Mg_{1/3}Nb_{2/3})_{0.7}Ti_{0.3}O_3$ (PMN-PT) (011)[19,20]. These studies showed a generation of intrinsic anisotropic strain across the entire substrate which altered the switching of the MTJ, but required 100s of volts[19,20] because of the large thickness of the piezoelectric layers.

The voltage required to generate a given strain scales linearly with the piezoelectric layer thickness. Hence small layer thicknesses are conducive to low switching voltages and low energy dissipation. Unfortunately, scaling piezoelectric layers thicknesses to sub-micrometer dimensions is very challenging and there is no assurance that such ultrathin layers will retain strong piezoelectricity. There are two-dimensional piezoelectric like $CrTe_2$ with layer thicknesses of few nm that might have been preferred to reduce the switching voltage to perhaps sub-mV levels, but their piezoelectric coefficients $d_{33}$ are also small (~17 pC/N)[21] which could negate the advantage of the small thickness and keep the switching voltage well above 1 V.

One way to overcome this problem of the voltage scaling with piezoelectric layer thickness is to employ a time-varying strain via a surface acoustic wave (SAW) launched in the piezoelectric layer[22-27]. The SAW is confined to the surface of the piezoelectric, regardless of the layer thickness, and the time varying voltage required to generate the SAW has no dependence on the piezoelectric layer thickness. This approach allows the use of thick piezoelectric layers without incurring a concomitant increase in the switching voltage and energy dissipation. SAW-induced magnetization dynamics have been examined in Co microbars[24], SAW-based magnetization switching from a single domain to a vortex state has been observed[25] and SAW-induced switching in dipole coupled Co nanomagnets to implement a Boolean inverter has also been reported[26]. Furthermore, it has been shown that application of a SAW signal can quasi-statically reduce the energy barrier within the soft layer of an MTJ to assist in STT switching[28]. The effect of a SAW on MTJ dynamics has also been examined recently[29]. However, it was found that the change in the coercivity of the soft layer of an MTJ is negligible, indicating that the effective magnetic field generated by a SAW is miniscule compared to the coercive field[29].

This motivates the use of a local gate to apply strain concentrated around the MTJ region as a promising method for achieving higher strain without the need to apply high voltages. Another advantage of in-plane strain over anisotropic strain is the possibility to deposit a thin layer of piezoelectric film on a traditional Si substrate, for compatibility with CMOS technology. This technique has been previously explored with MTJs deposited on PMN-PT and reduces voltages to hundreds of volts for strain-mediated switching[30,31]. However, these studies have mostly been done on micron-sized MTJ devices[30]. For high-density applications, nanoscale MTJs are necessary for their use in integrated circuits. Some efforts have been made to investigate electric field control on nanoscale MTJ devices[31]. Despite this downscaling, previous work on nanoscale MTJs still requires at least 100 V to observe strain altering the magnetization switching. Therefore, this motivates us to enhance the optimization of the local gate and its placements, aiming to maximize strain induced by low voltages. This, in turn, enables us to effectively control magnetic and electric properties of sub-micron CoFeB MTJs.

In this work, we demonstrate that by applying voltage through a local gate on a PMN-PT (001) substrate interfaced with the MTJ free layer, we can control the switching field and $TMR$ of nanoscale MTJs with record-low voltage. The device stack consists of a CoFeB magnetic free layer and an antiferromagnetic IrMn pinned CoFeB fixed layer. An elliptical MTJ with diameters 480 nm × 520 nm is patterned and centered around voltage pads distanced 48 μm apart. We show stable resistance states with consistent cycle-to-cycle switching characteristics both with and without application of the external gate voltage ($V_g$). Our results shows that $V_g$ as low as 5 V, which is equivalent to an electric field of 1.04 kV/cm, can both tune the switching field and enhance the $TMR$ of the MTJs. An even more modest $V_g$ of 200 mV still modulates the device switching field. These results are important next steps in achieving energy-efficient straintronic devices based on MTJs.

**Results and Discussion**

The structure of the MTJ studied is shown in Fig. 1(a), PMN-PT (001) substrate | Ta (5) | CoFeB (3) | MgO (2) | CoFeB (3) | IrMn (12) | Ru (7.5); numbers are in nm thickness. The stack was sputter deposited at NVE corporation at base pressure of $10^{-8}$ Torr and was post UHV annealed at 300 °C for 2 hours in presence of magnetic field of 400 mT, along the easy axis of the MTJ film. The in-plane magnetic hysteresis loop of the unpatterned MTJ film is measured using a vibrating sample magnetometer, shown in Fig. 1(b). Lower coercivity is expected of the bottom free layer, consisting of a 3 nm CoFeB layer, while the top pinned CoFeB layer with IrMn is expected to have higher coercivity. The exchange bias field of the stack is about 8 mT.

Consequently, when strain is applied to the MTJs, alterations in the magnetic properties of the less-coercive free layer are expected.

Figure 2(a) depicts a schematic of the MTJ device, including the $V_g$, fabricated using electron beam lithography and ion beam etching techniques. The patterned elliptical MTJ has dimensions of 480 nm × 520 nm, with its easy axis aligned along the y-axis. Voltage pads with voltage $V_g$ and ground are positioned at approximately 45° angle, located roughly 17 μm and 23 μm along the x-axis and y-axis, respectively, from the center of the ellipse. The distance between the voltage pads measures 48 μm, and the MTJ is centered between these pads. Top-down scanning electron microscopy image of the device is shown in Fig. 2(b).

We first study the switching mechanism of MTJ device 1 without any $V_g$. Prior to the measurements, the device is saturated with a high magnetic field along the easy axis of the MTJs, and then magnetic field ($H$) is swept along the easy axis of the ellipse starting from 0 mT. The two-point resistance ($R$) is measured using a sensing current of 10 μA in all measurements.

Initially, at $H = 0$ the device is in the $AP$ state owing to the dipole coupling between the hard and the soft layers, transitioning to $P$ state with an increase in the external field, shown by blue arrows labeled 1 in Fig. 3. Subsequently, when $H$ is decreased from positive field value to 0, the MTJ switched from $P \rightarrow AP$ state, again owing to dipole coupling between the hard and the soft layers, as indicated by the green arrows. Similar switching behaviors from $AP \rightarrow P$ and $P \rightarrow AP$, shown by red and yellow arrows, are observed when $H$ is applied in the opposite direction. This switching mechanism has been well studied in in-plane MTJs[32,33]. Device 1 shows $\Delta R = R_{AP} - R_P$ of approximately 16 Ω, which is relatively low compared to state-of-the-art MTJs[22], due to the relatively rough PMN-PT surface compared to a silicon substrate. Nevertheless, within the scope of our study, the observed stable *AP* and *P* states of the MTJ are adequate for investigating the effects of local strain on nanoscale MTJ devices.

We then study the switching behavior of device 1 by applying a small $V_g$ = 200 mV. Figure 4(a) shows the $\Delta R$ vs. $H$ loop of the MTJ at $V_g$ = 0 V and $V_g$ = 200 mV. At both voltages, the value of $\Delta R$ is measured to be 16 Ω. With regards to coercivity, the loop is significantly broader at $V_g$ = 200 mV on the negative side of the loop compared to $V_g$ = 0 V. After saturation in the negative direction, the switching field of the MTJ shifts from -3.8 mT at $V_g$ = 0 V to -1.12 mT at $V_g$ = 200 mV. This is in stark contrast to what has been reported in the case of time varying strain (or SAW), where there was almost no effect on the coercivity[29] because the time-averaged value of the strain in a sinusoidal SAW is zero. The $V_g$-mediated change in the switching field of the MTJ indicates an enhancement in magnetic anisotropy. The strain induced magnetic anisotropy $K_{me} = \frac{3}{2}\lambda_s \sigma$, and $\sigma = (\epsilon_{xx} - \epsilon_{yy})Y$ where $\epsilon_{xx}$ is strain along the x axis, $\epsilon_{yy}$ is strain along the y axis, and Y is Young's modulus. The value of $\lambda_s$ has been measured to be positive for CoFeB films[34]. Therefore, for $V_g$ = 200 mV, an enhancement in magnetic anisotropy along the y axis suggests that $\epsilon_{yy} > \epsilon_{xx}$. Figure 4(b) shows three sets of cycle-to-cycle data for switching the MTJ at both $V_g$ = 0 V and $V_g$ = 200 mV. All the blue curves exhibit broadened field loops, and the switching field from negative saturation has changed to -1.12 mT, which is consistent across all three measurements. This demonstrates the impact of a voltage as low as 200 mV, in contrast to previous studies that have shown high voltage requirement for the tunability of switching field and coercivity with voltage.

To further investigate the tunability of magnetoresistance with $V_g$, we study switching behavior of a second MTJ (device 2) that has the same dimensions as device 1. Figure 5(a) displays $\Delta R$ vs. $H$ for device 2 at various $V_g$ = 0 V, 5 V, 10 V and 20 V. Similar to device 1, device 2 starts at the *AP* state for $H = 0$. Sweeping $H$ in the positive direction, we observe switches at 8.7 mT, 7.1 mT, 8.2 mT and 7.8 mT, corresponding to voltages of 0 V, 5 V, 10 V and 20 V, respectively, showing no significant change in the switching field for device 2 from $AP \rightarrow P$. However, when $H$ decreases from positive to zero, there is a broadening of the field loop when $V_g$ is on. We surmise that this broadening of the field loop is due to enhancement in magnetic anisotropy along the y axis, similar to what was observed in device 1. As $V_g$ increases, $R_{AP}$ increases. $\Delta R$ increases by 20% at $V_g$ = 5 V, 72 % at 10 V and 86 % at 20 V as shown in Fig. 5(b). This is a very different behavior from what has been reported for the effect of strain on MTJs, where one usually observes a decrease in $\Delta R$ rather than an increase[35]. Strain modifies the crystal structure of the MgO tunnel barrier layer[36]. As the MgO structure gets strained, it modifies the conductance in both the parallel and the antiparallel configuration of the ferromagnetic electrode by moving minority states away from the $\Gamma$ point and, leading to a change in the $TMR$. Usually, the tunneling probability in both the parallel and antiparallel channels decrease under strain and the sign of the change in the $TMR$ is determined by which decreases more. In our case, the tunneling probability in the antiparallel channel seems to have decreased more, in contrast to the case reported previously[35].

Our results demonstrate that by optimizing the local gate placement, we can control the switching field and resistance state of MTJ devices. This local gating scheme allows for the generation of a tunable anisotropic in-plane strain within an isotropic piezoelectric material. In previous reports, the electric field required for magnetization rotation typically amounts to approximately 8 kV/cm, equivalent to a 400 V voltage applied across substrates with a thickness of 0.5 mm[37,38]. However, in these experiments, we observe a significant alteration in the switching field and $TMR$ within a voltage range of 200 mV to 20 V, corresponding to electric field range of 1.04 kV/cm to 4.17 kV/cm, due to proximity the voltage pads. We show that for nanoscale MTJ devices, voltage pads at roughly 24 μm can significantly reduce the operating $V_g$. Our results show that the voltage required for strain-enhanced switching in nano-scale devices is at least 20 times lower than that of devices without a local gate scheme and at least 5 times lower than that of state-of-art local gate-tuned straintronics devices. These findings also indicate that by maximizing strain through local gating, it is possible to avoid the need to integrate complex high-magnetostrictive materials into the MTJ thin film stack.

**Conclusions**

In summary, we have demonstrated electric field manipulation of nanoscale MTJs deposited on a PMN-PT (001) substrate by using local gate pads spaced 50 μm apart for strain generation. This strain is highly concentrated along the MTJ region, resulting in significant change in switching field and $TMR$ upon application of voltage ranging from 200 mV to 20 V. The applied voltage changes the switching field of the MTJ through enhancing the magnetic anisotropy, and it also linearly enhances the $TMR$ through straining the crystal structure of the tunnel barrier. These findings highlight the potential for electric field manipulation and strain engineering as effective strategies for controlling the properties and functionality of nanoscale MTJs. This ability to modulate switching behavior and $TMR$ through record-low voltages could lead to energy efficient MTJ-based straintronic devices.

**Acknowledgements**

The authors acknowledge funding support from the National Science Foundation Division of Computing and Communication Foundations (NSF CCF) under Grant Numbers 2006753 (S. K., J. A. I.) and 2006843 (S. B.). The authors acknowledge the use of shared research facilities supported in part by the Texas Materials Institute and the Texas Nanofabrication Facility supported by NSF Grant No. NNCI-1542159.

**Author Contributions**

S. Karki conducted the experiments and prepared the manuscript. J. Kwon, R. Fabiha, V. Rogers and T. Leonard helped with the experiments and manuscript preparation. J. Davies provided discussions and material stack for MTJ. S. Bandyopadhyay provided ideas and discussions of importance to the work. J. A. C. Incorvia created, led and supervised the work and manuscript.

**Data Availability**

The datasets generated during and/or analyzed during the current study are available from the corresponding author on reasonable request.

**Conflict of Interest**

On behalf of all authors, the corresponding author states that there is no conflict of interest.

**Figures**

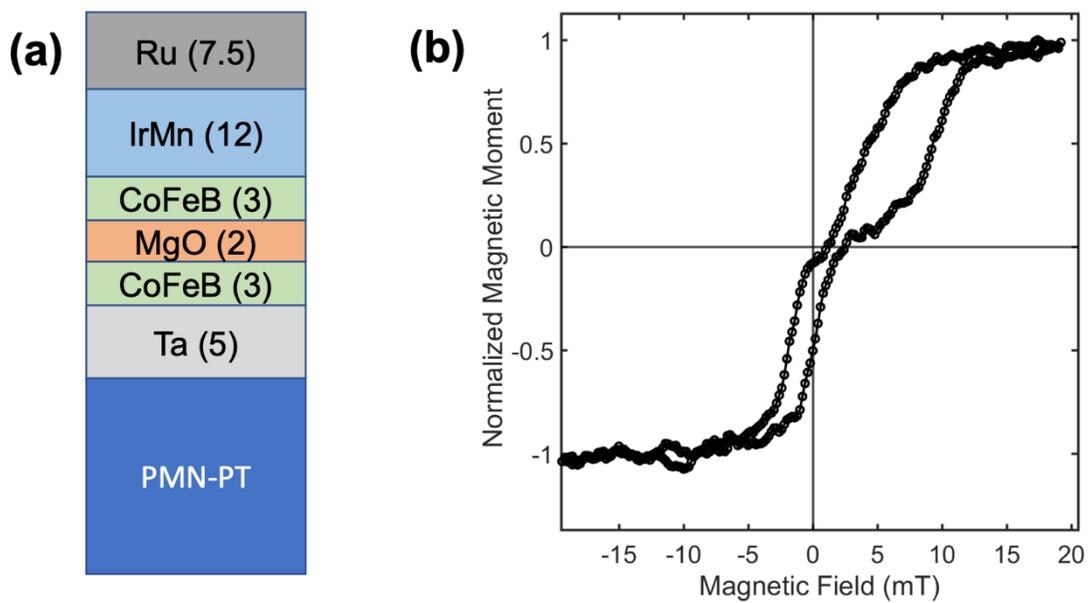

**Figure 1:** (a) Schematic of the MTJ stack with layers sputter deposited on a PMN-PT (001) substrate; numbers in parentheses are in nm. (b) In-plane magnetic hysteresis loop of the unpatterned MTJ stack.

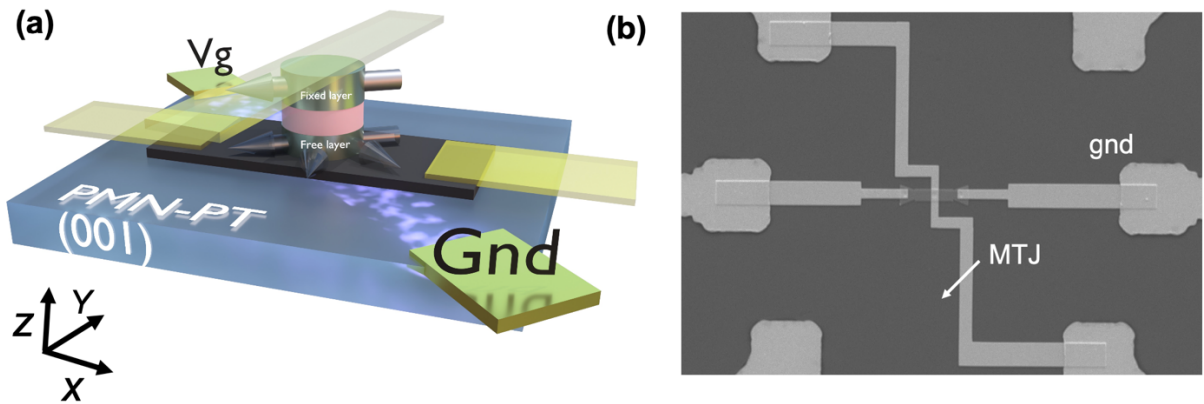

**Figure 2:** (a) Schematic of the MTJ device structure with local gate voltage ($V_g$) pads. (b) Top-down scanning electron microscope image of a 480 nm × 520 nm MTJ.

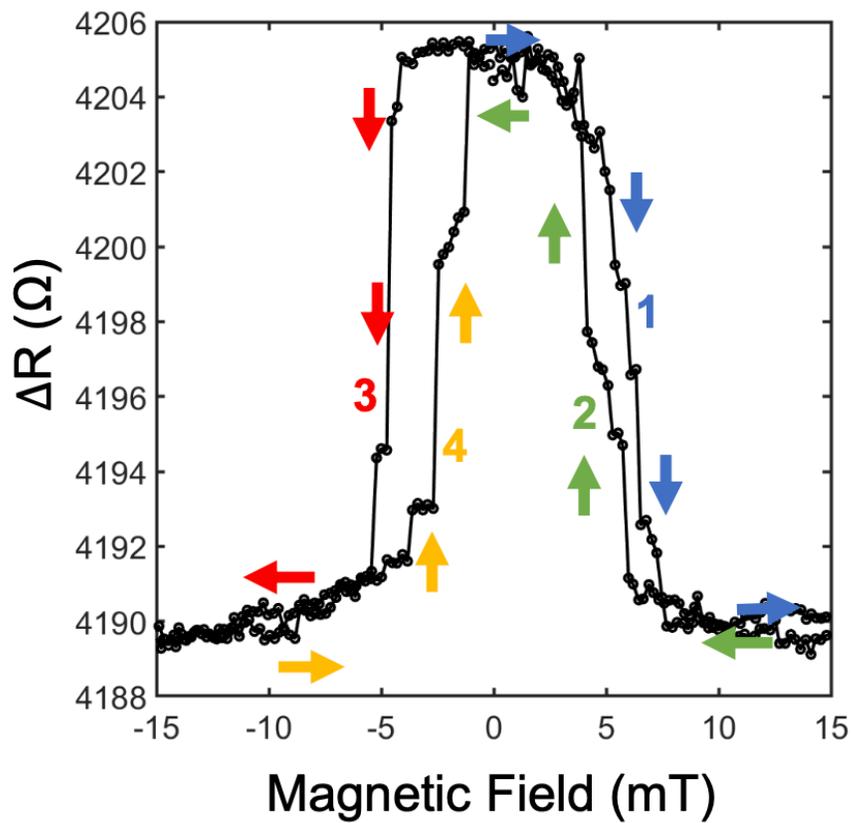

**Figure 3:** Field loop ($R$ vs. $H$) for the MTJ device.

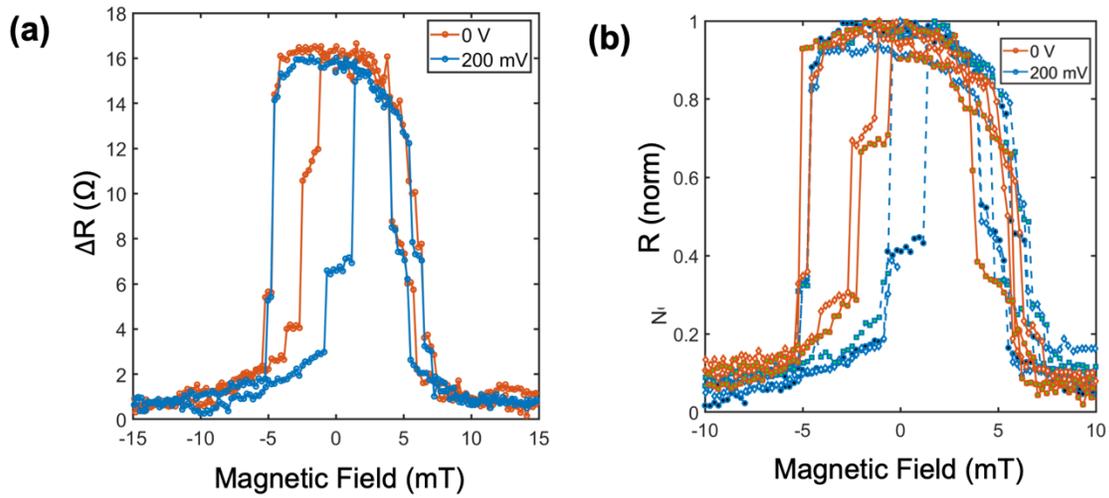

**Figure 4:** (a) Field loop (Δ$R$ vs. $H$) of a 480 nm × 520 nm elliptical MTJ at $V_g = 0$ V and $V_g = 200$ mV. (b) Cycle-to-cycle variation for $V_g = 0$ V and $V_g = 200$ mV.

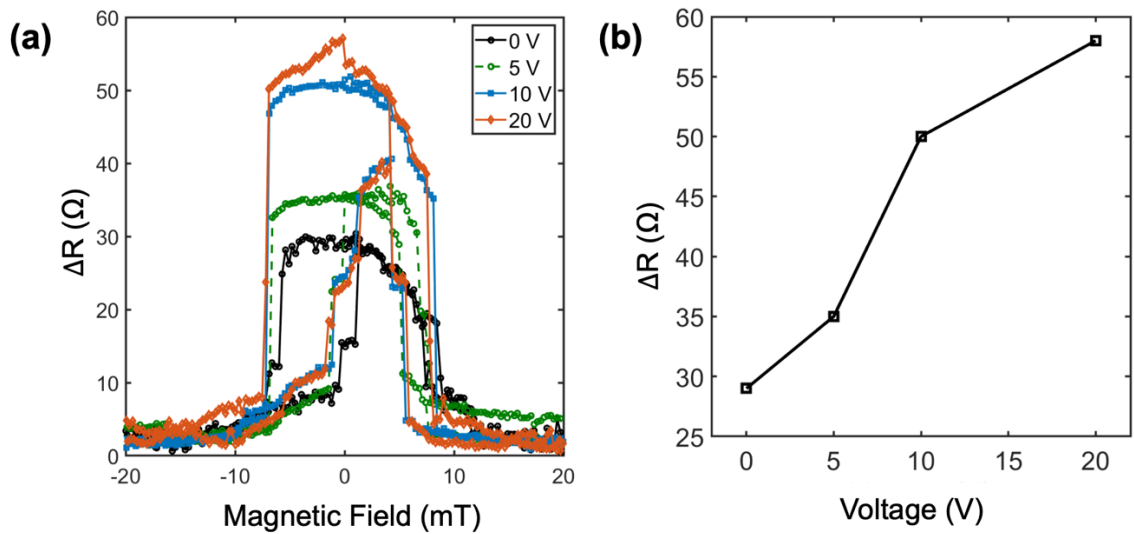

**Figure 5:** (a) Field loop (Δ$R$ vs. $H$) for the elliptical MTJ device 2 (480 nm × 520 nm) at $V_g = 0$ V, 5 V, 10 V, and 20 V. (b) Δ$R$ vs $V_g$.